\documentclass[aps,prd,twocolumn,superscriptaddress,amsfont,amssymb,amsmath,nofootinbib,showpacs,balancelastpage]{revtex4-1}
\usepackage[latin1]{inputenc}
\usepackage{amsmath}
\usepackage{amsfonts}
\usepackage{appendix}
\usepackage{amssymb}
\usepackage{epsfig}
\usepackage{graphicx}
\usepackage{dcolumn}
\usepackage{bm}
\usepackage{epstopdf}
\usepackage{color}
\usepackage[usenames,dvipsnames,svgnames]{xcolor}
\usepackage[colorlinks=true,
            linkcolor=red,
            urlcolor=gray,
            citecolor=blue]{hyperref}
\usepackage{multirow}
\usepackage{float}

\usepackage{subfigure}

\usepackage{enumitem}

\def\3nab{\tilde{\nabla}}

\def\be {\begin{equation}}
\def\ee {\end{equation}}
\def\ba {\begin{align}}
\def\ea {\end{align}}

\def\bc {\begin{center}}
\def\ec {\end{center}}
\def\case#1/#2{\frac{#1}{#2}}

\newcommand{\bver}{\begin{verbatim}}
\newcommand{\ever}{\end{verbatim}}

\newcommand{\bea}{\begin{eqnarray}}
\newcommand{\eea}{\end{eqnarray}}
\newcommand{\beaa}{\begin{eqnarray*}}
\newcommand{\eeaa}{\end{eqnarray*}}

\def\case#1/#2{\textstyle\frac{#1}{#2}}

\begin{document}
%%%%%%%%%%%%%%%%%%%%%%%%%%%%%%%%%%%%%%%%%%%%%%%%%%%%%%%%%%%
\title{Parametrizing growth in dark energy and modified gravity models}
%%%%%%%%%%%%%%%%%%%%%%%%%%%%%%%%%%%%%%%%%%%%%%%%%%%%%%%%%%%

\author{
Miguel Aparicio Resco 
}
\email{migueapa@ucm.es}
\affiliation{Departamento de F\'{\i}sica Te\'orica, Universidad Complutense de Madrid, 28040 
Madrid, Spain}
\author{Antonio L.\ Maroto}
\email{maroto@ucm.es}
\affiliation{Departamento de F\'{\i}sica Te\'orica, Universidad Complutense de Madrid, 28040 
Madrid, Spain}


%%%%%%%%%%%%%

%\date{\today}

\begin{abstract} 
It is well-known that an extremely accurate parametrization of the 
growth function of matter density perturbations in $\Lambda$CDM cosmology, with errors below $0.25 \%$,  
is given by  $f(a)=\Omega_{m}^{\gamma} \,(a)$ with  $\gamma \simeq 0.55$.  
In this work, we show that a simple modification of this expression
also provides a good description of growth in modified gravity theories. 
We consider the model-independent approach to modified  
 gravity in terms of an effective Newton constant written as 
 $\mu(a,k)=G_{eff}/G$ and show that $f(a)=\beta(a)\Omega_{m}^{\gamma} \,(a)$
 provides fits to the numerical solutions with similar accuracy to that of $\Lambda$CDM.
 In the time-independent case with $\mu=\mu(k)$, 
 simple analytic expressions for $\beta(\mu)$ and $\gamma(\mu)$ are presented.
 In the time-dependent (but scale-independent) case $\mu=\mu(a)$, we show
 that $\beta(a)$ has the same time dependence as $\mu(a)$. As an example, explicit formalae are provided in the DGP model.  In the general
 case, for theories with $\mu(a,k)$, we obtain a perturbative expansion for $\beta(\mu)$ around  the General Relativity case $\mu=1$  which,  for $f(R)$ theories, reaches an accuracy below $1 \%$. Finally, as an example we apply the obtained  fitting functions in order to forecast the precision with which future galaxy surveys will be able to measure the $\mu$ parameter. 
\end{abstract} 

%%%%%%%%%%%%%

\maketitle
%%%%%%%%%%%%%%%%%%%%%%%%%%%%%%%%%%%%%%%%%%%%%%%%%%%%%%%%%%%%%%%%%%%%%%%%%%%%%%%%%%%%
\section{Introduction}
%%%%%%%%%%%%%%%%%%%%%%%%%%%%%%%%%%%%%%%%%%%%%%%%%%%%%%%%%%%%%%%%%%%%%%%%%%%%%%%%%%%%

One of the most important open problems in cosmology is to understand the 
physics behind cosmic acceleration \cite{Riess:1998cb, Perlmutter:1998np, Eisenstein:2005su, Percival:2007yw}. Most of the models proposed to date can be classified in two major categories, namely, dark energy and modified gravity.
The former refers to a new component which acts as a source of gravity within 
the framework of General Relativity (GR), the simplest example 
being a cosmological constant. By modified gravity we understand 
extensions of GR which include new degrees of freedom that mediate the gravitational
interaction. Well-known examples of these theories are the $f(R)$ models \cite{Nojiri:2003ft, Capozziello:2003tk} or the Dvali, Gabadadze y Porrati model (DGP) \cite{Dvali:2000hr}.

The main source of observational information about cosmic acceleration comes
from distance measurements which can map the expansion history of the universe.
However, this kind of observations alone are not able to completely discriminate between 
the different theoretical approaches. In recent years the construction of 
large galaxy catalogues  has opened the possibility of mapping not 
only the expansion but also the growth history of large-scale structures. 
In general, modified gravity theories change the relation between the 
density perturbations and the gravitational potentials,  thus modifying the 
amplitude of matter density perturbations as a function of time (growth function).
They also modify the equations determining the form of the 
gravitational potential entering in the photon propagation equation (lensing potential). Thus, observations of clustering and lensing 
at different redshifts provide a way to break the degeneracy between different acceleration models.

For the sake of concreteness, we will consider scalar perturbations around a flat FLRW background
in the longitudinal gauge \cite{Tsujikawa},
\begin{eqnarray}
ds^2=-(1+2\Psi)dt^2+a^2(t)(1+2\Phi)d{\bf x}^2
\end{eqnarray}
At the background level, the model is characterized by 
its Hubble function $H(a)=\dot a/a$ which allows to determine 
the cosmological distances.

At the perturbation level, 
the modified Einstein equations in a wide range of gravity theories, 
for sub-Hubble scales and in the quasi-static approximation 
lead to a modified Poisson equation that in Fourier space reads
\begin{eqnarray}
k^2\Psi\simeq -4\pi G_{\text{eff}}\,a^2\rho_m\delta_m
\end{eqnarray}
where $\delta_m=\delta \rho_m/\rho_m$ is the matter density contrast which is related to the galaxy density contrast $\delta_g$ by the bias factor $b$ as $\delta_g=b\,\delta_m$.  Here $G_{\text{eff}}$ is the effective gravitational coupling which will differ in general 
from the bare gravitational constant $G=1/(8\pi M_p^2)$. The 
growth equation reads
 \begin{eqnarray}
\delta''_m+\left(2+\frac{H'}{H}\right)\delta'_m-\frac{3}{2}\mu(a,k)\Omega_m(a)\delta_m\simeq 0 \label{growth}
\end{eqnarray}
where prime denotes derivative with respect to $\ln a$ and $\Omega_{m} (a)$ is the matter density parameter $\Omega_{m} (a)=\Omega_m \, a^{-3} \, \frac{H_{0}^{2}}{H^{2}(a)}$.

\begin{figure*} [htb]
 \begin{center}
 \hspace*{-1.8cm}\includegraphics[width=1.2\textwidth]{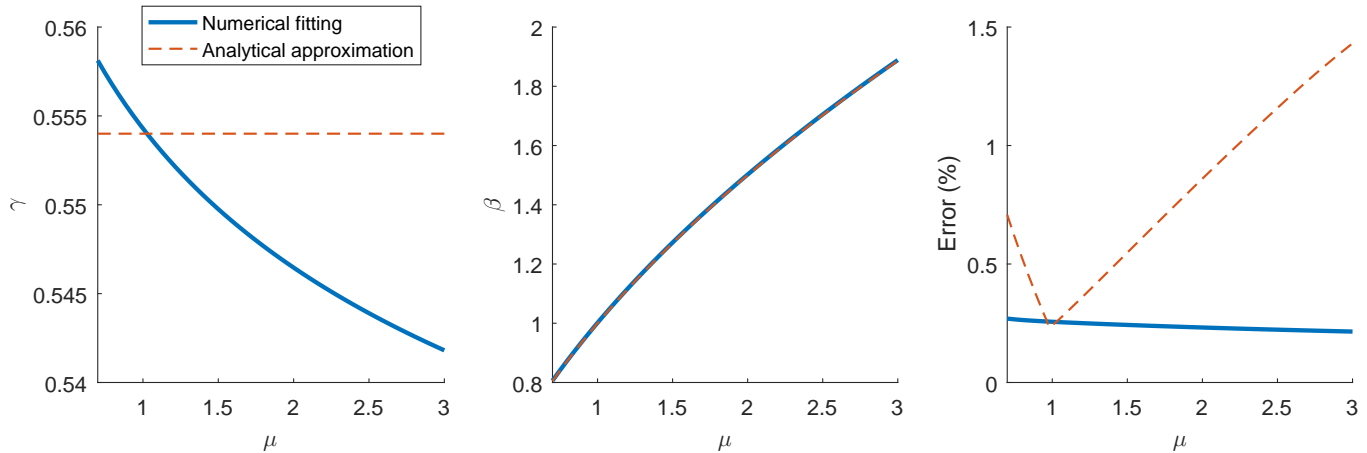}
  \end{center}
  \caption{From left to right: $\gamma$, $\beta$ and relative error in the growth function as a function of $\mu$. The dashed red line corresponds to analytical approximation in (\ref{15}) with $\gamma=\gamma_*$ and $\beta$ in (\ref{14a}). The continuous blue line, corresponds to the numerical fitting of $\beta$ and $\gamma$
  to expression (\ref{15a}).}
 \label{Figure_1a} 
\end{figure*}
On the other hand, the lensing potential satisfies
\begin{eqnarray}
\frac{\Psi-\Phi}{2}\simeq -\frac{3G_{\text{eff}}}{2G}\frac{1+\eta}{2}\left(\frac{aH}{k}\right)^2\Omega_m(a)\delta_m \label{lensing}
\end{eqnarray}
These modifications of Einstein equations are parametrized by two free
functions:
\begin{eqnarray}
\mu(a,k)=\frac{G_{\text{eff}}}{G}
\end{eqnarray}
and 
\begin{eqnarray}
\eta(a,k)=-\frac{\Phi}{\Psi}\, .
\end{eqnarray}
For any local and generally covariant four-dimensional theory of 
gravity, it can be shown that they reduce to rational functions of $k$
which are even in theories with purely scalar extra degrees of 
freedom. If we also assume
that no higher than second derivatives appear in the equations of motions, then they can be completely described by five functions of time only  as follows \cite{Silvestri:2013ne}
\begin{eqnarray}
\mu(a,k)=\frac{1 + p_3(a)k^2}{p_4(a) + p_5(a)k^2}\label{mu}
\end{eqnarray}
and 
\begin{eqnarray}
\eta(a,k)=\frac{p_1(a) + p_2(a)k^2}{1 + p_3(a)k^2}\, .
\end{eqnarray}
Thus, deviations from $\mu=\eta=1$ signal a breakdown of standard GR.

Once the evolution of perturbations have been solved, the growth function is defined as,
\begin{equation}\label{4}
f(a)=\frac{d\, \ln(\delta_m)}{d\, \ln(a)}.
\end{equation}
In the case of $\Lambda$CDM, a good approximation for the growth function is given by,
\begin{equation}\label{5}
f(a)=\Omega_{m}^{\gamma}(a),
\end{equation}
where  $\gamma$ is known as the growth index which has a value in $\Lambda$CDM of $\gamma_*\simeq0.55$ \cite{Linder:2005,Linder:2007hg}. This expression provides
accuracies better than $0.25 \%$, and accordingly could be useful in the data analysis of 
present and future surveys such as J-PAS \cite{J-PAS}, DESI \cite{DESI} or 
Euclid \cite{Euclid} which will be able to measure $f(z)$ with precisions 
around $1\%-3\%$.

In the modified gravity case, (\ref{5}) does not necessarily  provide 
a good  fit and different alternatives have been considered in the literature, mainly focused on the modification of the growth index. 
Such alternative expressions have been obtained on a case by case basis and to 
the best of our knowledge no model-independent analysis has been performed so far. 
Thus the aim of this work is to fill this gap and determine
accurate fitting functions for the modified growth function in terms of the $\mu$ parameter
in a model-independent way.

The paper is organized as follows: in section \ref{sec1} we analyze the case in which $\mu \, (a,k)$ does not depend on redshift, and we find an approximate analytic expression for the growth function. In section  \ref{sec2} we consider the time-dependent but scale-independent case and apply our results to the DGP model. In  \ref{sec1a}, we propose a perturbative parametrization for the  general case $\mu \, (a,k)$ and consider $f(R)$ theories as an example. 
In section \ref{sec7}, we make use of the obtained fitting functions in order to
forecast the precision with which future galaxy surveys will be 
able to measure the $\mu$ parameter in the redshift-independent case.
Finally in section \ref{sec6} we briefly discuss the results and conclusions.

\begin{figure*}[htb]
\centering
  \begin{tabular}{@{}cc@{}}
    \includegraphics[width=.4975\textwidth]{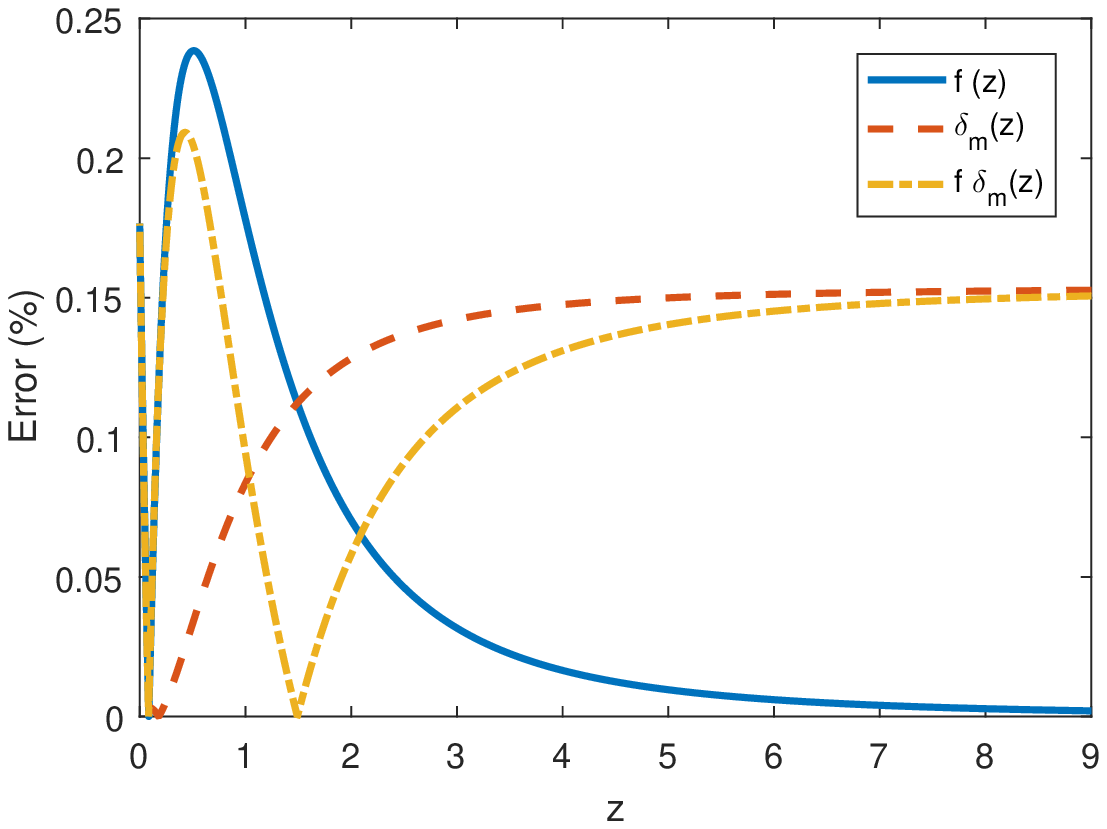} &
    \includegraphics[width=.4975\textwidth]{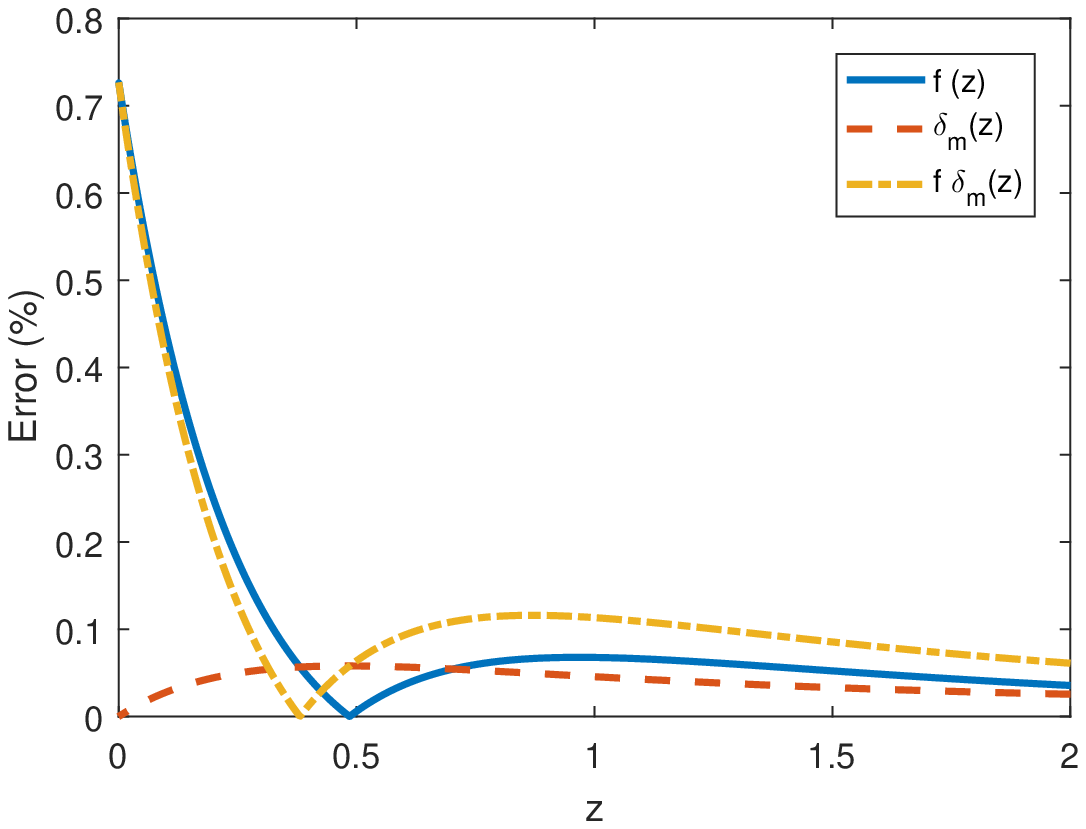} 
  \end{tabular}
  \caption{Errors for $f(z)$, $\delta_m (z)$ and $f(z) \delta_m (z)$ in $\Lambda$CDM using $\gamma_*=0.554$ (left) and $\gamma_*=0.550$ (right). We see that for $\gamma_*=0.554$ the error in $f(z)$ is below $0.25 \%$ but that in $\delta_m(z)$
  reaches $0.15\%$, whereas for $\gamma_*=0.550$, the error in $\delta_m(z)$ can be reduced below $0.05\%$, however
  the corresponding error for $f(z)$ grows to $0.7\%$.}
  \label{Figure_1b}
\end{figure*}

%%%%%%%%%%%%%%%%%%%%%%%%%%%%%%%%%%%%%%%%%%%%%%%%%%%%%%%%%%%%%%%%%%%%%%%%%%%%%%%%%%%%
\section{Time-independent case $\mu=\mu(k)$}\label{sec1}
%%%%%%%%%%%%%%%%%%%%%%%%%%%%%%%%%%%%%%%%%%%%%%%%%%%%%%%%%%%%%%%%%%%%%%%%%%%%%%%%%%%%

We start by studying the simplest case in which  functions $p_{i} (a)=p_i$ in (\ref{mu}) are constant so that $\mu \, (a,\,k)=\mu(k)$ does not depend on the scale factor. Although in general for modified gravities with extra scalar or
vector degrees of freedom we expect both time and scale dependence of 
the $\mu$ factor, the time-independent case can be used for phenomenological
parametrizations of the effective Newton constant at different length scales.
It can also be applied in scalar-tensor models as \cite{Gannouji:2008jr} in which $G_{eff}$ rapidly tends to a constant at high redshift.

 In this simple case, we can develop an analytical study. Making the change of variable from $\delta_m$ to $f$ in equation (\ref{growth}) we have,
\begin{equation}\label{7}
f'+f^{2}+\left(2+\frac{H'}{H}\right) \, f-\frac{3}{2} \, \mu \, (k) \, \Omega_{m} (a)=0 .
\end{equation}
Let us first assume for simplicity a $\Lambda$CDM  background. Later on we will consider a more general case with time-dependent effective equation of state for the dark energy or modified gravity extra components. The Hubble parameter in $\Lambda$CDM reads, 
\begin{equation}\label{8}
H(a)=H_{0}\,\sqrt{\Omega_{m} \, a^{-3}+\Omega_{\Lambda}},
\end{equation}
with this expression and taking into account the definition of $\Omega_{m} (a)$ in function of $H$, we can obtain,
\begin{equation}\label{9}
\frac{H'}{H}=-\frac{3}{2}\,\Omega_{m} (a),
\end{equation}
and
\begin{equation}\label{10}
\Omega_{m}'(a)=-3\,\Omega_{m}(a)\,(1-\Omega_{m}(a)).
\end{equation}
Inspecting the numerical solutions of (\ref{7}), we can see that a parametrization that provides a good fit is of the form,
\begin{equation}\label{11}
f(a,k)=\beta(k) \, \Omega_{m}^{\gamma(k)}(a),
\end{equation}
where for each Fourier mode, $\beta(k)$ and $\gamma(k)$ are in general the constants to adjust. This parametrization type was considered in \cite{DiPorto:2007ovd} for the particular case of a scalar-tensor theories. Using equations (\ref{9}) and (\ref{10}) together with (\ref{11}) in equation (\ref{7}) we have,
\begin{eqnarray}\label{12}
(2-3\gamma) \, \Omega_{m}^{\gamma-1}(a)+\left(3\gamma-\frac{3}{2}\right) \, \Omega_{m}^{\gamma}(a)\nonumber\\
+\beta \, \Omega_{m}^{2\gamma-1}(a)-\frac{3}{2} \, \frac{\mu}{\beta}=0.
\end{eqnarray}

Although this expression can be satisfied exactly only in the case in which $\Omega_{m}(a)=\Omega_m$ is a constant, as commented before,  it is possible to obtain approximate solutions in the general case. Thus, for example, substituting $\mu=\beta=1$, we recover the case of $\Lambda$CDM so that for $\gamma=\gamma_*$  equation (\ref{12}) is satisfied so that,
\begin{eqnarray}\label{13}
(2-3\gamma_*) \, \Omega_{m}^{\gamma_*-1}(a)+(3\gamma_*-3/2) \, \Omega_{m}^{\gamma_*}(a)\nonumber\\
+\Omega_{m}^{2\gamma_*-1}(a)-\frac{3}{2}\simeq 0,
\end{eqnarray}
Thus, if we take $\gamma=\gamma_*$ in the case with $\mu$ and $\beta$  
different from one, we get from (\ref{12}) and (\ref{13})
\begin{eqnarray}\label{14}
3\left(1-\frac{\mu}{\beta}\right)+2(\beta-1) \, \Omega_{m}^{2\gamma_*-1}(a)\simeq 0. 
\end{eqnarray}
We consider a final approximation. Since $\gamma_*\simeq 1/2$, we assume that $\Omega_{m}^{2\gamma_*-1}(a)\simeq1$. Thus we are able to find a relationship between $\mu$ and $\beta$,
%
%\begin{eqnarray}\label{14a}
%\eta=\frac{5}{4}\, \left[ \sqrt{1+\frac{24}{25} \, \alpha}-1 \right]. 
%\end{eqnarray}
\begin{eqnarray}\label{14a}
\beta=\frac{1}{4}\, \left[ \sqrt{1+24 \, \mu}-1 \right]. 
\end{eqnarray}

Therefore, we have an analytic expression which is an approximate solution of  equation (\ref{7}) for redshift-independent $\mu(k)$
\begin{equation}\label{15}
f(a,k)=\frac{1}{4}\, \left[ \sqrt{1+24 \, \mu(k)}-1 \right] \, \Omega_{m}^{\gamma_*}(a).
\end{equation}

The error of this approximation generally depends on the scale factor and reaches a 
maximum at $a=1$ as discussed below. In Fig. \ref{Figure_1a}, we plot the maximum  error as a function of $\mu$ for  $\Omega_{m}=0.271$. We have taken this particular value in order to compare with previous works although we have checked that
the results remain unchanged in the range $\Omega_m=0.27-0.31$ which includes the 
latest  Planck value \cite{Planck}. The error corresponds to the difference between the fitting function and the numerical solution of (\ref{7}) divided by their average value. We can see that the error is always below $2 \%$. Notice that  we have considered a wide range of $\mu$ values, although relatively small deviations from $\mu=1$ could generate a large integrated Sachs-Wolfe effect.

\begin{figure*} [htb]
 \begin{center}
 \hspace*{-1.5cm}\includegraphics[width=1.1\textwidth]{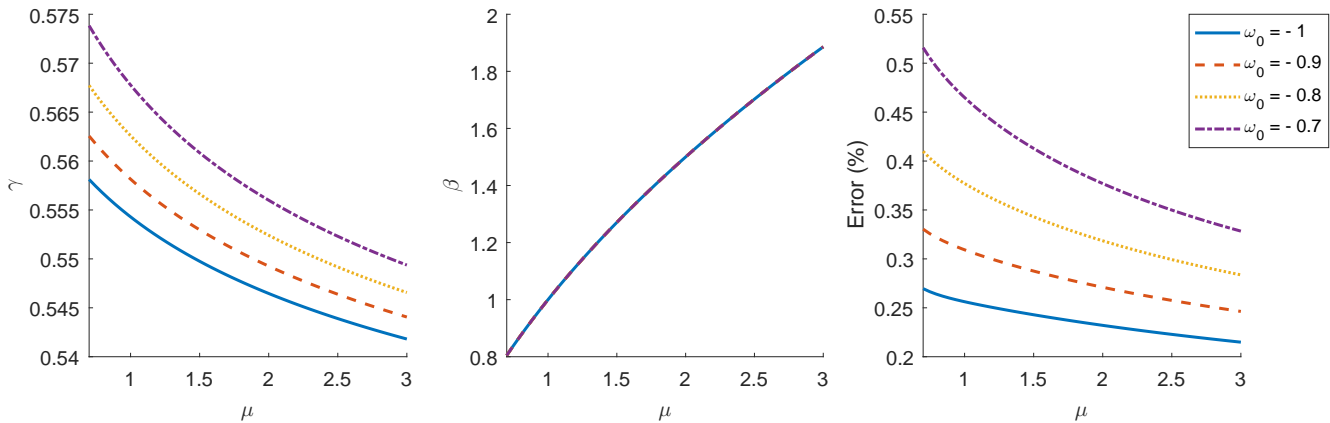} 
  \end{center}
  \caption{From left to right,  functions $\gamma \,(\mu)$ (\ref{eqg}), $\beta\, (\mu)$ (\ref{14a}) and growth function relative error, for time-independent $\mu$  for different values of $\omega_{0}$ with $\omega_{1}=0$. We have assumed that $\Omega_{m}=0.271$.}
 \label{Figure_2} 
 \end{figure*}
\begin{figure*} [htb]
 \begin{center}
 \hspace*{-1.5cm}\includegraphics[width=1.1\textwidth]{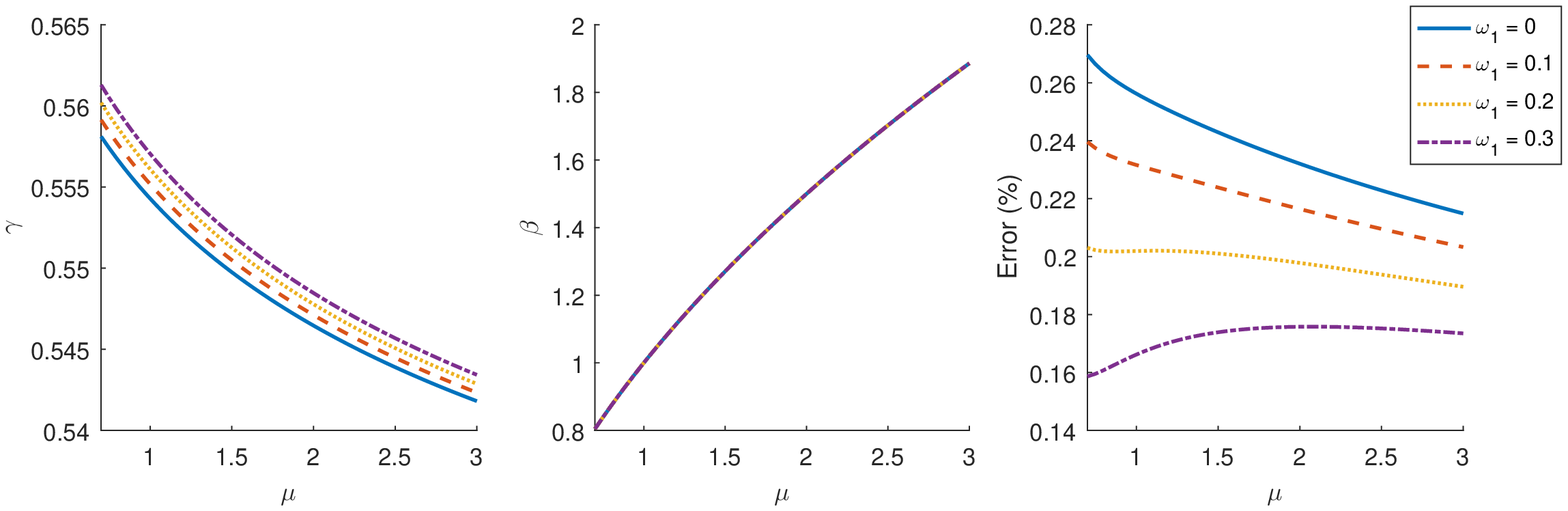} 
  \end{center}
  \caption{From left to right,  functions $\gamma \,(\mu)$ (\ref{eqg}), $\beta\, (\mu)$ (\ref{14a}) and growth function relative error, for time-independent $\mu$  for different values of $\omega_{1}$ with $\omega_{0}=-1$. We have assumed that $\Omega_{m}=0.271$.}
 \label{Figure_3} 
 \end{figure*}

It is however possible to improve the fit if we allow also the growth index $\gamma$ to depend on $\mu$. In this case, it is possible to find a good agreement with  
\begin{equation}\label{17a}
\gamma \,(\mu)=\frac{1}{2}+\frac{0.161}{1.967+\beta \, (\mu)},
\end{equation}
with $\beta(\mu)$ given in (\ref{14a}). In Fig. \ref{Figure_1a} we plot this relation together with the error 
corresponding to the improved growth function
\begin{equation}\label{15a}
f(a,k)=\frac{1}{4}\, \left[ \sqrt{1+24 \, \mu(k)}-1 \right] \, \Omega_{m}^{\gamma(\mu(k))}(a).
\end{equation}
We can see that in this case, the maximum error can be below $0.25 \%$. Also, we see that for GR  ($\mu=1$), we obtain $\gamma_* = 0.554$ as the value for the best fit, 
which is slightly different from that quoted in \cite{Linder:2005} ($\gamma_*=0.550$).
In order to understand the difference, we analyze  the error in three different  functions, $f(z)$, $\delta_m (z)$ and $f(z) \delta_m (z)$ (see also \cite{Linder2}). As we can see in Fig. \ref{Figure_1b}, the error in $f(z)$ is larger for low redshift than the error in $\delta_m (z)$, but in general the errors for the three functions are of the same order. Notice that the value $\gamma_*=0.554$  minimizes the error in $f(z)$, however, the error in $\delta_m(z)$
is minimized by $\gamma_*=0.550$ \cite{Linder:2005}. This value also 
minimizes the error both in $f(z)$ and $f(z)\delta_m(z)$ for $z>0.4$ Since the error in 
the observable $f(z)\sigma_8(z)$ is dominated by the error in $f(z)$ we have taken $\gamma_*=0.554$ as our reference value in this work. 

On the other hand,  at early times, in the matter dominated era, 
$\Omega_{m}(a)\simeq 1$ and equation (\ref{7}) can be solved exactly
\begin{equation}\label{17}
f(a,k)=\frac{1}{4}\, \left[ \sqrt{1+24 \, \mu(k)}-1 \right],
\end{equation}  
therefore $f(a,k)$ is just constant in time. The fitting function (\ref{15}) exactly agrees with this result for $\Omega_m(a)=1$ and this is the reason why the error increases as we move away from the matter era. Thus for matter domination the density contrast grows as 
\begin{equation}\label{18}
\delta_m(a,k)\propto a^{\frac{1}{4}\, \left[ \sqrt{1+24 \, \mu(k)}-1 \right]}, 
\end{equation}  
in agreement with \cite{Fry,Gannouji:2008jr}. This implies that if we want to preserve the growth of density contrast proportional to the scale factor in the matter era, $\mu$ should depend on the scale factor and tend to unity at early times.

\begin{figure} %[h!]
  	\includegraphics[width=0.4975\textwidth]{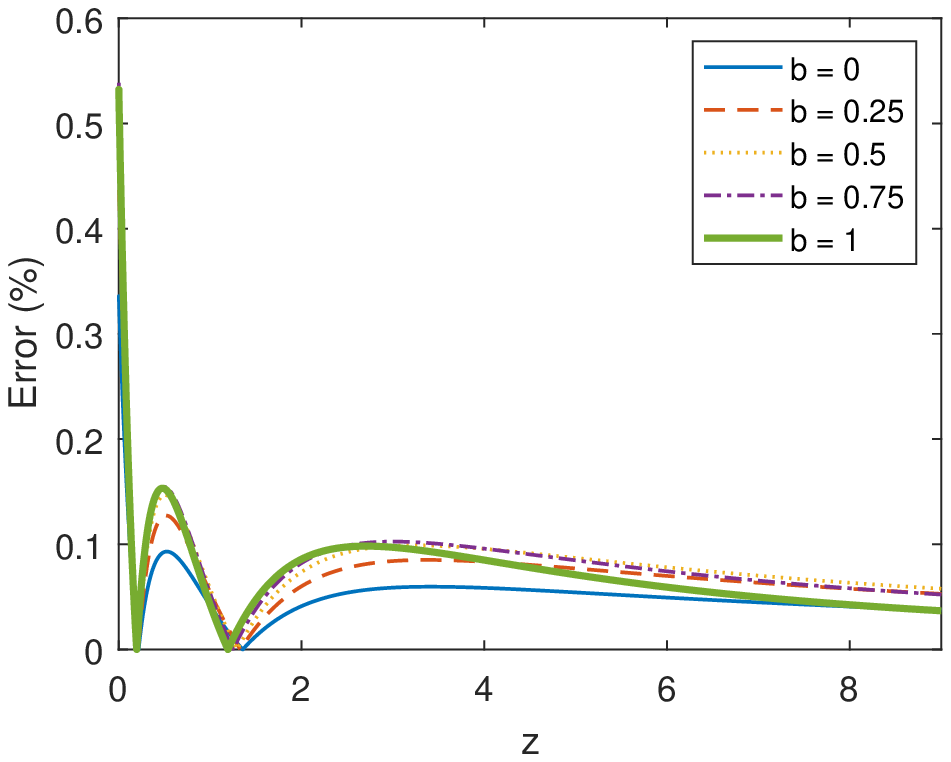}
		\caption{\footnotesize{Relative difference between the numerical solution and the fitting function (\ref{23}) for the model $\mu \, (a)=1+b \, a$. As we see, the maximum error is reached at $z=0$.}}
  \label{Figure_6}
\end{figure}
%%%%%%%%%%%%%%%%%%%%%%%%%%%%%%%%%%%%%%%%%%%
\subsection{Beyond the $\Lambda$CDM background}
%%%%%%%%%%%%%%%%%%%%%%%%%%%%%%%%%%%%%%%%%%

We now consider modifications of the background expansion. 
In order to keep the approach model-independent, we will parametrize them  
with an extra component $\Omega_{DE}(a)$ with arbitrary equation of state
$\omega_{DE}(a)$. This extra component could correspond
directly to dark energy or to the effective fluid description of the  modified gravity. 
At late times, i.e. neglecting the radiation contribution,  
we can write
\begin{equation}\label{19}
H(a)=H_{0}\,\sqrt{\Omega_{m} \, a^{-3}+\Omega_{DE} (a)},
\end{equation}
with 
\begin{eqnarray}
\Omega_{DE}(a)=\Omega_{DE}\,\exp\left(\int_0^a\frac{3(1+\omega_{DE}(\hat a))}{\hat a}d\hat a\right).
\end{eqnarray}
Using (\ref{19}) to obtain $\Omega_m(a)$, we get the expressions that replace (\ref{9}) and (\ref{10}),
\begin{equation}\label{20}
\frac{H'}{H}=-\frac{3}{2}\,\Omega_{m} (a)+\frac{\Omega_{m}(a)}{2 \, \Omega_{m}} \, a^{3} \, \Omega_{DE}'(a),
\end{equation}
and
\begin{equation}\label{21}
\Omega_{m}'(a)=-3\,\Omega_{m}(a)\,(1-\Omega_{m}(a))-\frac{\Omega_{m}^{2}(a)}{\Omega_{m}}  \, a^{3} \, \Omega_{DE}'(a).
\end{equation}
Following the same procedure as above, we obtain the analogous equation to (\ref{12}) with an extra term, 
\begin{eqnarray}\label{22}
(2&-&3\gamma) \, \Omega_{m}^{\gamma-1}(a)+\left(3\gamma-\frac{3}{2}\right) \, \Omega_{m}^{\gamma}(a)\nonumber\\
&+&\beta \, \Omega_{m}^{2\gamma-1}(a)-\frac{3}{2} \, \frac{\mu}{\beta}\nonumber\\
&+&\left(\frac{1}{2}-\gamma\right)\frac{\Omega_{DE}'(a)}{\Omega_{m}}\, a^{3} \, \Omega_{m}^{\gamma}(a)=0.
\end{eqnarray}
We see that the new term is proportional to $(\frac{1}{2}-\gamma)$, so it is expected that it does not increase the errors in an important manner. 
Thus, considering the fitting function with $\gamma=\gamma_*$ in (\ref{15}), we can see, that the errors increase in comparison with those for the $\Lambda$CDM background  up to $3 \%$.

As we did in the $\Lambda$CDM case, we can obtain better fits by modifying  
the expressions for $\beta(k)$ and $\gamma(k)$. Thus, the analysis shows that 
$\beta(k)$ is not sensitive to $\omega_{DE}(a)$ and 
therefore (\ref{14a}) provides a good approximation also in this case. The 
expression for $\gamma$ is however modified. 

For example for the effective equation of state given by \cite{Chevallier:2000qy,Linder:2002et},
\begin{equation}\label{3}
\omega_{DE}(a)=\omega_{0}+\omega_{1} \, (1-a).
\end{equation}
with $\omega_0$ and $\omega_1$ constants, we find
\begin{equation}\label{eqg}
\gamma (\mu, \omega_{0}, \omega_{1})=\frac{1}{2}+\frac{0.161}{\beta(\mu)-1.967\omega_{0}-0.4789\omega_{1}}
\end{equation}

Thus, the growth function reads in this case
\begin{equation}
f(a,k)=\frac{1}{4}\, \left[ \sqrt{1+24 \, \mu(k)}-1 \right] \, \Omega_{m}^{\gamma(\mu(k),\omega_{0}, \omega_{1})}(a).
\end{equation}

In  Fig. \ref{Figure_2} and Fig. \ref{Figure_3} we can see that the error can be reduced to $0.5\%$  with 
this parametrization.

\begin{figure} [htb]
  	\includegraphics[width=0.4975\textwidth]{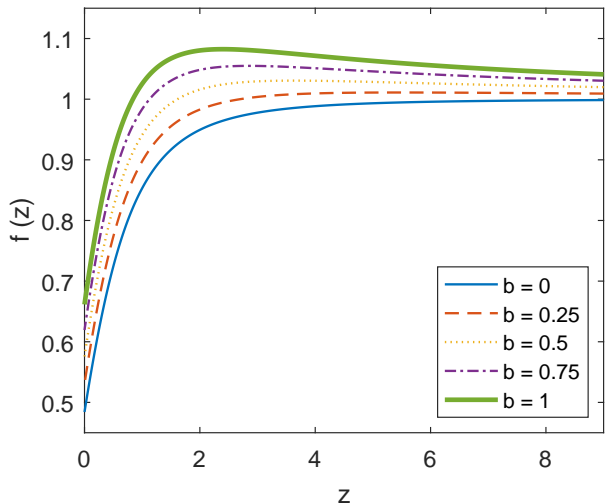}
		\caption{\footnotesize{Numerical solutions in the model $\mu \, (a)=1+b \, a$ for different values of $b$. The fit according to the equation (\ref{23}) is not represented because it differs in less than $0.6 \, \%$ with respect to the numerical solution.}}
  \label{Figure_7}
\end{figure}
\begin{figure*} %[h!]
 %\begin{center}
 \hspace*{-1.8cm}\includegraphics[width=1.2\textwidth]{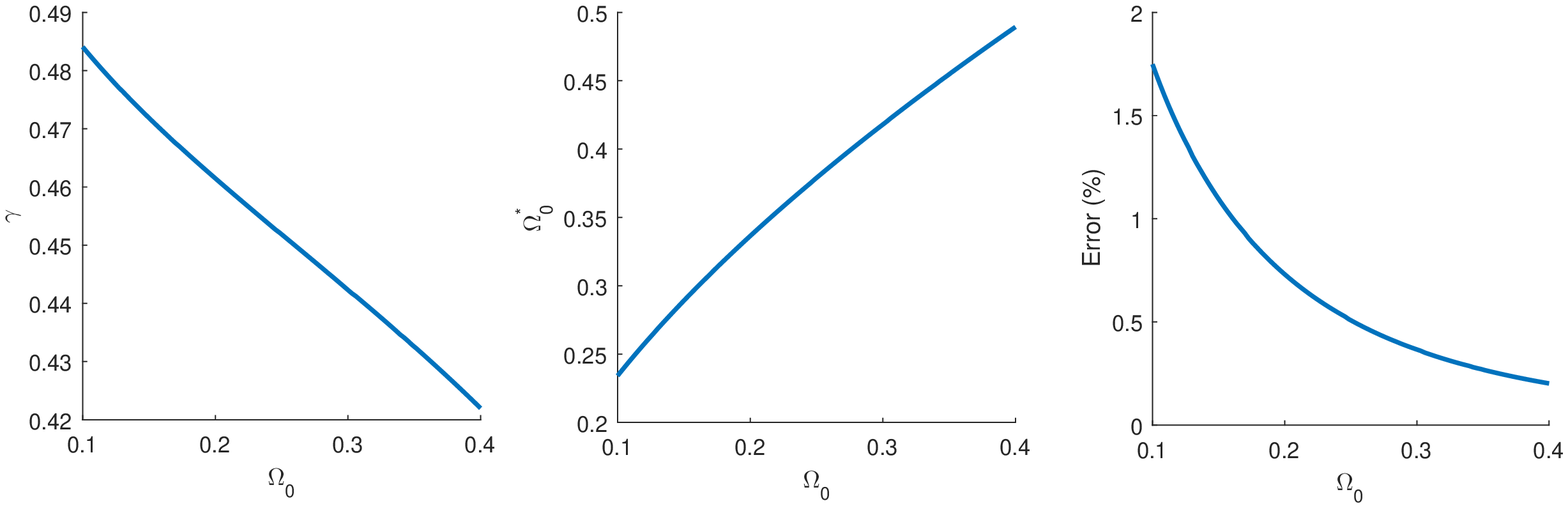} 
  %\end{center}
  \caption{From left to right, the functions $ \gamma \, (\Omega_ {0}) $, $ \Omega_ {0} ^ {*} \, (\Omega_ {0}) $ and the relative error of the growth function for the DGP model. The error decreases as we increase the value of $ \Omega_ {0} $. The maximum error is less than $ 0.5\, \% $ for typical values of $\Omega_ {0}$.}
 \label{Figure_8} 
 \end{figure*}

%%%%%%%%%%%%%%%%%%%%%%%%%%%%%%%%%%%%%%%%%%%%%%%%%%%%%%%%%%%%%%%%%%%%%%%%%%%%%%%%%%%%
\section{scale-independent case $\mu=\mu(a)$}\label{sec2}
%%%%%%%%%%%%%%%%%%%%%%%%%%%%%%%%%%%%%%%%%%%%%%%%%%%%%%%%%%%%%%%%%%%%%%%%%%%%%%%%%%%%
We have just seen that in the time-independent $\mu$ case,  an ansatz of the form,
\begin{equation}\label{fsi}
f(a)=\beta\;\Omega_{m}^{\gamma}(a),
\end{equation}
provides a good fit to the numerical solutions. Let us now consider the case in which $\mu=\mu \, (a; p_{1},..., p_{n})$ where $[p_{1},..., p_{n}]$ are the set of $n$ cosmological parameters that $\mu$ depends on, i.e. $\mu$  can depend on redshift(time) but not on the scale $k$. We will explore a similar ansatz for the growth function,
\begin{equation}\label{23}
f(a)=\beta \, (a; p^{*}_{1},..., p^{*}_{n}) \, \Omega_{m}^{\gamma}(a),
\end{equation}
where, in general, $p^{*}_{i}=p^{*}_{i}(p_{1},..., p_{n})$, $\gamma=\gamma(p_{1},..., p_{n})$ and let us assume that the $\beta$ function has the same $a$-dependence as the $\mu$ function (see \cite{Dent:2009wi, BuenoSanchez:2010wd} for similar proposals in particular models). 

Thus, let us consider a simple example.  For instance if  $\mu \, (a)=1+b \, a$ being $b$ dimensionless constant, then we consider $\beta=1+b^*a$, with $b^*=b^*(b)$. 

In Fig. \ref{Figure_6} we show the fit error  for different values of $b$. In Fig. \ref{Figure_7} we can see the growth functions for the corresponding values of $b$.
A $\Lambda$CDM background with $\Omega_{m}=0.271$ has been assumed. 
 It can be seen how it grows with redshift reaching values larger than one and then decreases tending to one when matter starts dominating. We can see that in this simple example, the parametrization (\ref{23}) provides fitting errors below $0.5 \%$ in the whole redshift range, but they are even below $0.2 \%$ for $z>0.1$ .

After we have studied this simple model, and checked the usefulness of parametrization (\ref{23}), we will apply it to more realistic models of modified gravity, such as DGP and certain phenomenological parametrizations of $\mu \, (a)$.

\begin{figure*} %[h!]
 \begin{center}
 \includegraphics[width=1\textwidth]{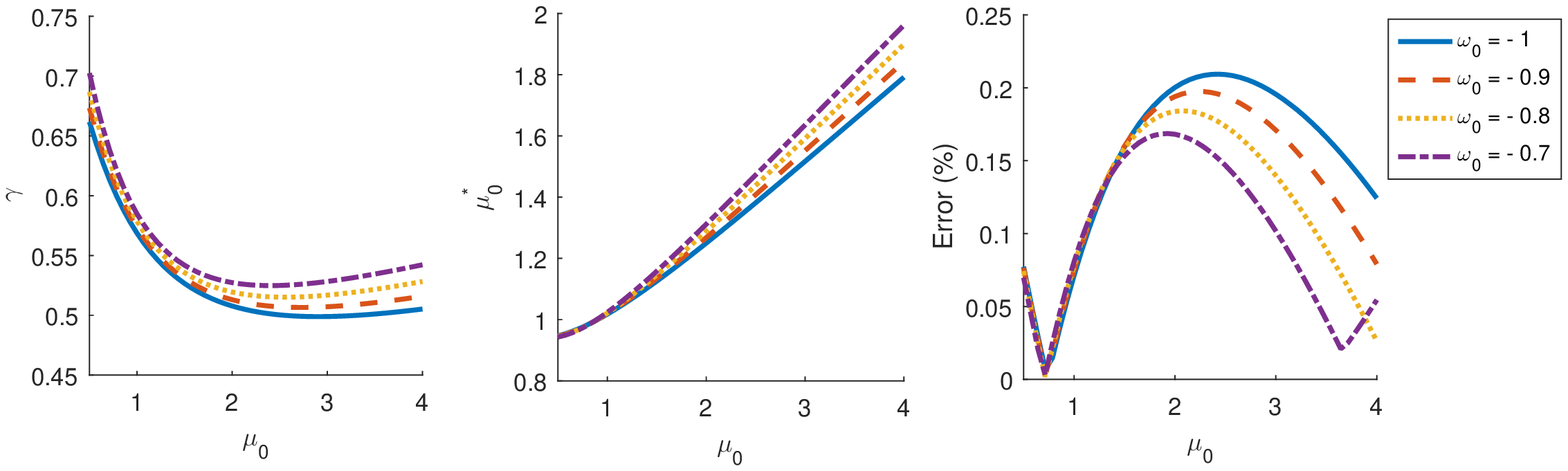} 
  \end{center}
  \caption{From left to right, the functions $\gamma\, (\mu_{0})$, $\mu_{0}^{*}\, (\mu_{0})$ and relative error of the growth function for the phenomenological  model (\ref{32}) for different values of $\omega_{0}$, setting $\omega_{1}=0$. The case with $\omega_{0}=-1$ is parametrized in (\ref{emu}) and (\ref{ecg}).}
 \label{Figure_11} 
 \end{figure*}
\begin{figure*} %[h!]
 \begin{center}
 \hspace*{-0.5cm}\includegraphics[width=1.07\textwidth]{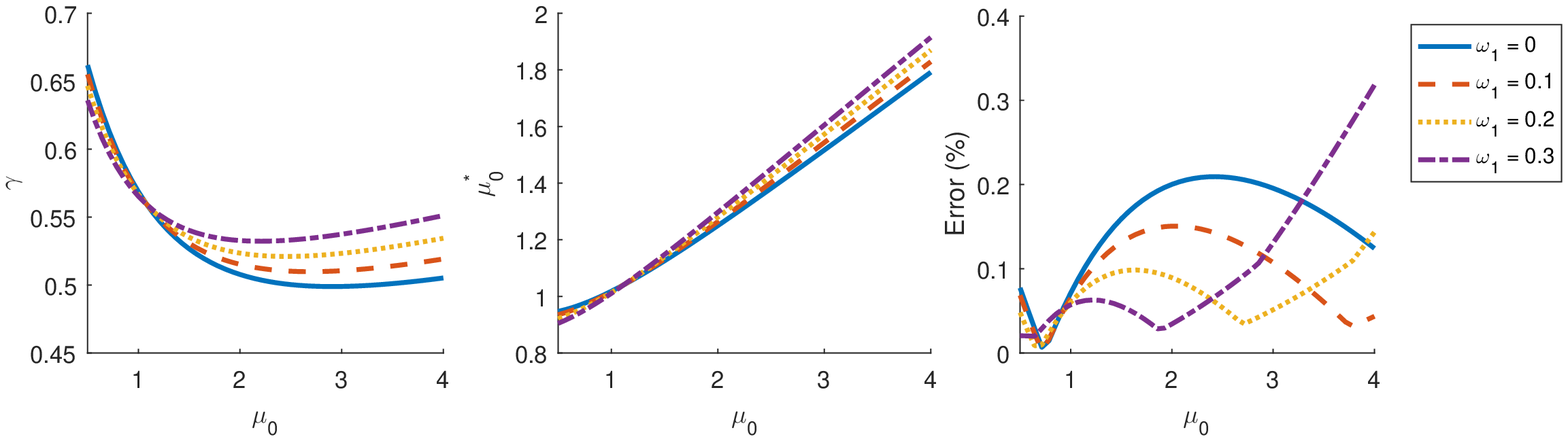}  
  \end{center}
  \caption{From left to right, the functions $\gamma\, (\mu_{0})$, $\mu_{0}^{*}\, (\mu_{0})$ and relative error of the growth function for the phenomenological  model (\ref{32}) for different values of $\omega_{1}$, setting $\omega_{0}=-1$. The case with $\omega_{1}=0$ is parametrized in (\ref{emu}) and (\ref{ecg}).}
 \label{Figure_12} 
\end{figure*}
%
%%%%%%%%%%%%%%%%%%%%%%%%%%%%%%%%%%%%%%%%%%%%%%%%%%%%%%%%%%%%%%%%%%%%%%%%%%%%%%%%%%%%
\subsection{DGP Model}\label{sec3}
%%%%%%%%%%%%%%%%%%%%%%%%%%%%%%%%%%%%%%%%%%%%%%%%%%%%%%%%%%%%%%%%%%%%%%%%%%%%%%%%%%%%  

In the DGP model \cite{Dvali:2000hr}, the background evolution differs from $\Lambda$CDM so that
\begin{equation}\label{24}
\Omega_{m} \, (a)=\frac{\Omega_{0} \, a^{-3}}{[(1-\Omega_{0})/2+\sqrt{\Omega_{0} \, a^{-3}+(1-\Omega_{0})^{2}/4}]^{2}},
\end{equation}
and
\begin{equation}\label{25}
\frac{H'}{H}=-\frac{3 \, \Omega_{m}(a)}{1+\Omega_{m}(a)}.
\end{equation}
being $\Omega_0=\Omega_m(a=1)$ the only free parameter at the background level.
The modified gravity parameters in this case read \cite{Kunz}
\begin{equation}\label{26}
\mu\, (a)=\frac{2(1+2\,\Omega_{m}^{2} \, (a))}{3(1+\Omega_{m}^{2} \, (a))} 
\end{equation}
and 
\begin{equation}
\eta\, (a)=\frac{2+\Omega_{m}^{2}(a)}{1+2\,\Omega_{m}^{2}(a)}.
\end{equation}
which are both $k$-independent, so that we try the parametrization in  (\ref{23}).

Thus, using (\ref{24}), (\ref{25}) and (\ref{26}) in equation (\ref{growth}), solving numerically and fitting to (\ref{23}), we get
\begin{equation}\label{27}
f(a)=\frac{2(1+2\,\Omega_{m}^{2} \, (a;\,\Omega^{*}_{0}))}{3(1+\Omega_{m}^{2} \, (a;\,\Omega^{*}_{0}))} \, \Omega_{m}^{\gamma(\Omega_0)}(a),
\end{equation}
where $\Omega_{m} \, (a;\,\Omega^{*}_{0})$ follows equation (\ref{24}) replacing $\Omega_{0}$ by $\Omega^{*}_{0}$. So we have two parameters to fit $\Omega^{*}_{0}\,(\Omega_{0})$ and $\gamma\,(\Omega_{0})$, which are given by the following expressions
\begin{eqnarray}
\Omega^*_0(\Omega_0)=0.8\,\Omega_0^{0.536}
\end{eqnarray}
and
\begin{eqnarray}
\gamma(\Omega_0)=0.52-0.47\,\Omega_0+\Omega_0^2-1.2\,\Omega_0^3
\end{eqnarray}

In Fig. \ref{Figure_8} we plot the maximum error for $f(z)$ which is always 
reached at $z=0$ as a function of $\Omega_0$. We see that for typical values of $\Omega_0\simeq 0.3$, the maximum error is below $0.5\,\%$.

Other procedure used to parametrize growth in DGP is to assume that the growth index depends on redshift. Thus, for example, from  equation (\ref{7}), $\gamma \, (z)$ has been obtained to first order in $(1-\Omega_{m} \, (a))$ in \cite{Ishak:2009qs}. Another possibility is to use  a parametrization like,
\begin{equation}\label{24a}
\gamma \,(z)=\gamma_{0}+\gamma_{1} \, \frac{z}{1+z}.
\end{equation}
Adding more terms, and therefore more fixing constants,  a reduced error was obtained
in \cite{Chen:2009ak}. See also \cite{Gong:2008fh}, for a different
parametrization. In the best cases, these methods reach errors similar to those
obtained in the present work.

%%%%%%%%%%%%%%%%%%%%%%%%%%%%%%%%%%%%%%%%%%%%%%%%%%%%%%%%%%%%%%%%%%%%%%%%%%%%%%%%%%%%
\subsection{Phenomenological parametrizations}\label{sec5}
%%%%%%%%%%%%%%%%%%%%%%%%%%%%%%%%%%%%%%%%%%%%%%%%%%%%%%%%%%%%%%%%%%%%%%%%%%%%%%%%%%%% 
As a second example, we will study the parametrization for $\mu \, (a)$ introduced in \cite{Simpson:2012ra} and also considered in \cite{PlanckDE},
\begin{equation}\label{32}
\mu \, (a)=1+(\mu_{0}-1) \, \frac{1-\Omega_{m}\, (a)}{1-\Omega_{m}}.
\end{equation}
Let us consider once more the effective equation of state (\ref{3}), thus,
\begin{equation}\label{33}
\Omega_{m}\, (a)=\frac{\Omega_{m}}{\Omega_{m}+(1-\Omega_{m}) \, a^{-3\,(\omega_{0}+\omega_{1})} \, e^{-3 \, \omega_1 \,(1-a)}}, 
\end{equation}
where we have fixed $\Omega_{m}=0.271$. Thus following (\ref{23}), the growth function becomes,
\begin{equation}\label{34}
f(a)=\left(1+(\mu_{0}^{*}-1) \, \frac{1-\Omega_{m}\, (a)}{1-\Omega_{m}}\right) \, \Omega_{m}^{\gamma}\, (a),
\end{equation}

In this case, we only need to fit the parameters $\gamma =\gamma \, (\mu_{0})$ and $\mu_{0}^{*}=\mu_{0}^{*}\,(\mu_{0})$. We plot in Fig. \ref{Figure_11} functions $\mu_{0}^{*}\,(\mu_{0})$, $\gamma\,(\mu_{0})$ along with errors in $f(a)$, for different values of $\omega_{0}$, setting $\omega_{1}=0$. In Fig. \ref{Figure_12} the same functions are shown, in this case varying $\omega_{1}$ with $\omega_{0}=-1$. In the $\omega_0=-1$ and $\omega_1=0$ case, i.e. $\Lambda$CDM background, 
the fitting functions read
\begin{eqnarray}\label{emu}
\mu_0^*=0.961-0.132\,\mu_0+0.245\,\mu_0^2-0.066\,\mu_0^3+0.0065\,\mu_0^4\nonumber \\
\end{eqnarray}
and
\begin{eqnarray}\label{ecg}
\gamma=0.456+0.012\,\mu_0+0.403\,e^{-1.37\,\mu_0}\,.
\end{eqnarray}

We can see that the error is less than $0.25 \, \%$ even changing the effective equation of state. In this case, a fit of the form (\ref{24a}) does not reproduce well the numerical results.
\begin{figure*} [htb]
 \begin{center}
 \includegraphics[width=0.48\textwidth]{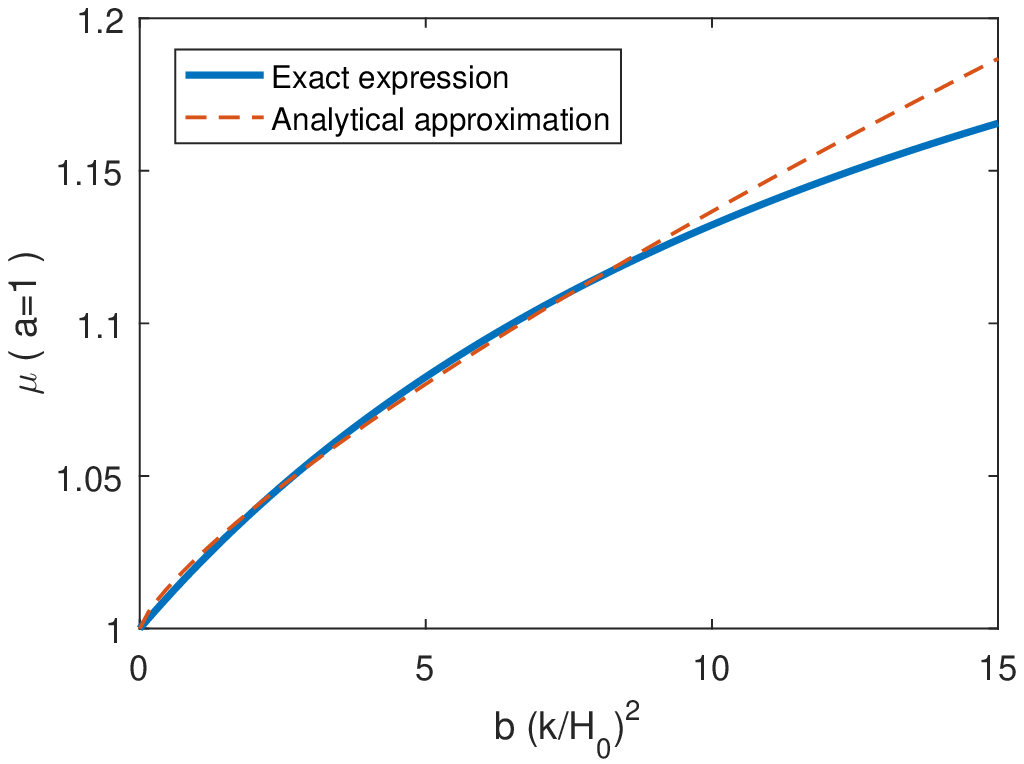}
 \includegraphics[width=0.48\textwidth]{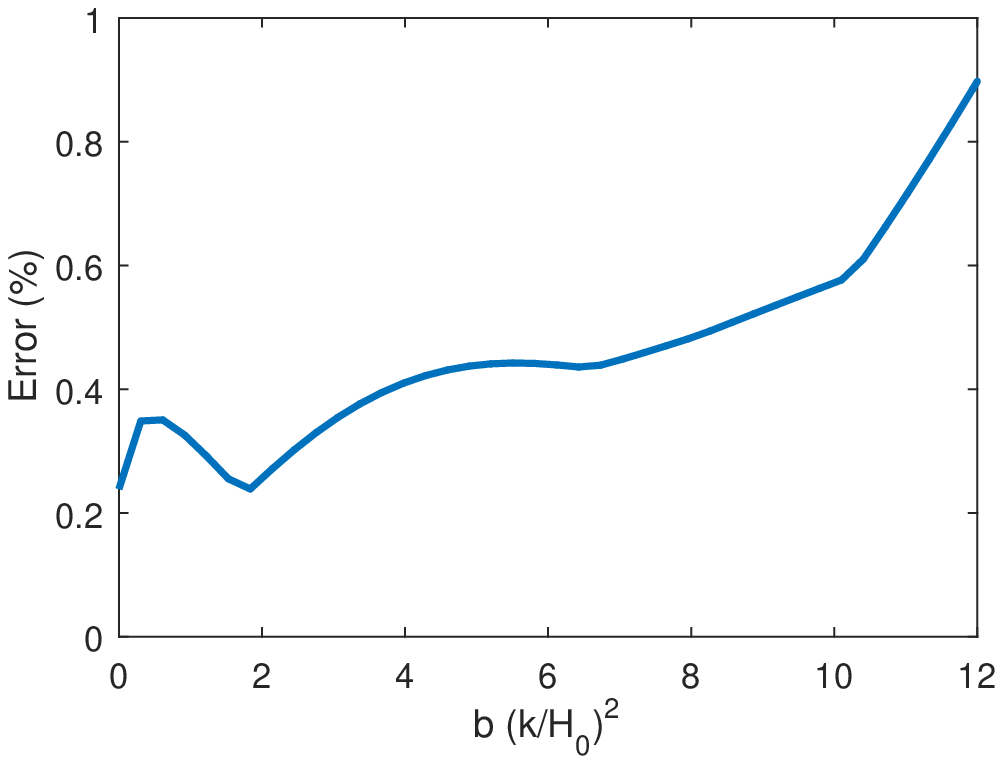}
  \end{center}
  \caption{Left panel: $\mu(a,k)$ with $a=1$ as a function of $b \, (k/H_0)^2$ for the $f(R)$ theory in (\ref{28d}). We compare the exact expression in (\ref{28a}) 
  with the approximation in equation (\ref{30d}). Right panel: the growth function relative error for the approximation (\ref{30d}) for the $\Lambda$CDM background.}
 \label{Figure_9} 
 \end{figure*}
%

%%%%%%%%%%%%%%%%%%%%%%%%%%%%%%%%%%%%%%%%%%%%%%%%%%%%%%%%%%%%%%%%%%%%%%%%%%%%%%%%%%%%
\section{General case: perturbative analysis}\label{sec1a}
%%%%%%%%%%%%%%%%%%%%%%%%%%%%%%%%%%%%%%%%%%%%%%%%%%%%%%%%%%%%%%%%%%%%%%%%%%%%%%%%%%%%
Let us consider the general case in which $\mu=\mu(a,\, k)$. In order to get closed expressions for the growth function, we will restrict ourselves to small perturbations around 
the GR case $\mu=1$.  
Thus, we start with equation (\ref{7}) and write 
\begin{eqnarray}
\mu(a,\,k) = 1+\alpha (a, \,k)
\end{eqnarray}
being $\vert\alpha\vert \ll 1$. Let us now write the perturbed growth function in the following form,
\begin{equation}\label{22a}
f(a,k) = [1+\epsilon \, (a, \,k)] \, f_{0}(a),
\end{equation}
with $\vert\epsilon\vert \ll 1$ and $f_{0}(a) = \Omega_{m}^{\gamma_*}(a)$ the 
growth function in $\Lambda$CDM. We insert (\ref{22a}) into (\ref{7}) and using that $f_{0}$ satisfies equation (\ref{7}) with $\mu = 1$, we obtain,
\begin{equation}\label{23a}
\epsilon'+\left[ \frac{3}{2} \Omega_{m}^{1-{\gamma_*}} + \Omega_{m}^{\gamma_*} \right] \, \epsilon = \frac{3}{2} \, \Omega_{m}^{1-{\gamma_*}} \, \alpha,
\end{equation}
where again, prime denotes derivative with respect to $\ln a$.  We can solve this equation analytically for a given initial condition at $a=a_i$ well inside the matter era where $\alpha(a_i,k) \simeq 0$ and $\epsilon(a_i,k) \simeq 0$, so that 
\begin{equation}\label{24b}
\epsilon \, (a,\,k) = \frac{3}{2} \,\, e^{-g(a)} \, \int_{a_{i}}^{a} \, \Omega_{m}^{1-{\gamma_*}}(a') \, \alpha  (a',\, k) \,\, e^{\,g(a')} \, \frac{da'}{a'},
\end{equation}
being,
\begin{equation}\label{25a}
g(a) = \int_{a_{i}}^{a} \, \left[ \frac{3}{2} \, \Omega_{m}^{1-{\gamma_*}}(a')+\Omega_{m}^{\gamma_*}(a') \right] \, \frac{da'}{a'}.
\end{equation}
The result does not depend on the particular value chosen (for concreteness, we took $a_i=10^{-2}$).

If we consider as a hard approximation that $\Omega_{m} \simeq 1$ then we can simplify equation (\ref{24b}) as,
\begin{equation}\label{26a}
\epsilon \, (a,\,k) = \frac{3}{2} \,\, a^{-\frac{5}{2}}  \, \int_{a_{i}}^{a} \, \alpha  (a',\, k) \,\, a'^{\frac{3}{2}}  \, da',
\end{equation}
and we can integrate by parts obtaining,
\begin{equation}\label{27a}
\epsilon \, (a,\,k) = \frac{3}{5} \, \sum_{n=0}^{\infty} \, \left( - \frac{2}{5} \right)^{n} \, \alpha^{(n)}(a,\, k),
\end{equation}
where $\alpha^{(n)}$ is the $n$-th derivative with respect to $\ln a$. Then, if we take $\alpha$ as a constant in $a$, $\epsilon \, (a,\,k) = \frac{3}{5} \alpha(k)$ and we recover the time-independent case that we analyzed above in (\ref{17}). 

In the following, we apply these results to different examples of modified gravity theories.

%%%%%%%%%%%%%%%%%%%%%%%%%%%%%%%%%%%%%%%%%%%%%%%%%%%%%%%%%%%%%%%%%%%%%%%%%%%%%%%%%%%%
\subsection{$f(R)$ Model}\label{subsec1}
%%%%%%%%%%%%%%%%%%%%%%%%%%%%%%%%%%%%%%%%%%%%%%%%%%%%%%%%%%%%%%%%%%%%%%%%%%%%%%%%%%%%
Let us consider $f(R)$ gravities \cite{Starobinsky:2007hu, Hu:2007nk, Appleby:2007vb, Tsujikawa:2007xu,  Tsujikawa:2009ku}. The growth function in this kind of  models has been studied in several works \cite{Gannouji:2008wt, Narikawa:2009ux,LopezRevelles:2013vm,Bamba:2012qi}. In particular, we will consider here the Hu-Sawicki model \cite{Hu:2007nk} written in the simple form,
\begin{equation}\label{28d}
f(R) = R-\frac{2 \Lambda}{1+\frac{b \, \Lambda}{R}},
\end{equation}
which, for $b=0$, reduces to the standard $\Lambda$CDM model with a cosmological constant $\Lambda$. The corresponding $\mu$ function reads \cite{Tsujikawa:2009ku}

\begin{equation}\label{28a}
\mu \, (a,k)=\frac{1}{f_{,R}}\frac{1+4(f_{,RR}(R_0)/f_{,R}(R_0))(k/a)^2}{1+3(f_{,RR}(R_0)/f_{,R}(R_0))(k/a)^2},
\end{equation}
where $f_{,R}$ and $f_{,RR}$ denote the first and second derivatives 
with respect to $R$ and $R_{0}$ is the scalar curvature assuming that the background agrees with that of $\Lambda$CDM
\begin{equation}\label{28b}
R_{0}(a)=3\,H_{0}^{2}\, [\Omega_{m} \, a^{-3}+4\, (1-\Omega_{m})].
\end{equation}
For small enough $b$, we can approximate,
\begin{equation}\label{aprox2}
\frac{f_{,RR}(R_0)}{f_{,R}(R_0)}\left(\frac{k}{a}\right)^2 \simeq \frac{4\,b}{3} \frac{(1-\Omega_{m})^{2}}{a^{2}r^{2}(a)} \frac{k^{2}}{H_{0}^{2}},
\end{equation}
where $r(a)=R_0(a)/3H_{0}^{2}$. Also, using that for small $x$,
\begin{equation}\label{aprox1}
\frac{1+4\, x}{1+3\,x} \simeq 1+0.44 \, x^{0.77},
\end{equation}
we can finally approximate (\ref{28a}) by
\begin{equation}\label{30d}
\mu(a,k) \simeq 1+0.44 \, \left[ \frac{4\,b}{3} \frac{(1-\Omega_{m})^{2}}{a^{2}r^{2}(a)} \frac{k^{2}}{H_{0}^{2}} \right]^{0.77},
\end{equation}
which allows to extract the explicit $b$ and $k/H_0$ dependence from the integral in equation (\ref{24b}). We compare in the left panel of Fig. \ref{Figure_9} this approximation with the exact expression. We see that it provides an 
excellent fit for $b(k/H_0)^2<10$. Then, using (\ref{24b}) we get
\begin{equation}\label{31d}
\epsilon(a,k)=\left[ b \, \left( \frac{k}{H_{0}} \right)^{2} \right]^{0.77} \, F(a),
\end{equation}
being,
\begin{align}\label{32d}
F(a)=0.8236 \, (1-\Omega_{m})^{1.54} \, e^{-g(a)} \, \nonumber \\
\times\int_{a_{i}}^{a} \, \frac{\Omega_{m}^{1-\gamma^*}(a')}{r^{2.31}(a')a'^{2.54}} \, e^{\, g(a')} \, da'.
\end{align}
The fitting functions for $F(a)$ in terms of $\Omega_m(a)$ can be easily obtained and reads
\begin{align}\label{33d}
F(a)=&0.140 \, \Omega_m(a)-0.545 \, \Omega_m^2(a)+0.994 \, \Omega_m^3(a) \nonumber\\
&-0.905 \, \Omega_m^4(a)+0.315 \, \Omega_m^5(a).
\end{align}
On the right panel of Fig. \ref{Figure_9}, we plot the growth function errors as a function of $b \, (k/H_0)^2$ using the expressions above. We see that the agreement with the numerical solution is better than $1\%$  when $\vert\mu-1\vert < 0.12$. Since these fits have been obtained 
for a $\Lambda$CDM  background, using the above expressions with different backgrounds would increase the errors up to $2\%$.

%%%%%%%%%%%%%%%%%%%%%%%%%%%%%%%%%%%%%%%%%%%%%%%%%%%%%%%%%%%%%%%%%%%%%%%%%%%%%%%%%%%%
\subsection{Phenomenological parametrization}\label{subsec3}
\begin{figure} %[h!]
  	\includegraphics[width=0.4975\textwidth]{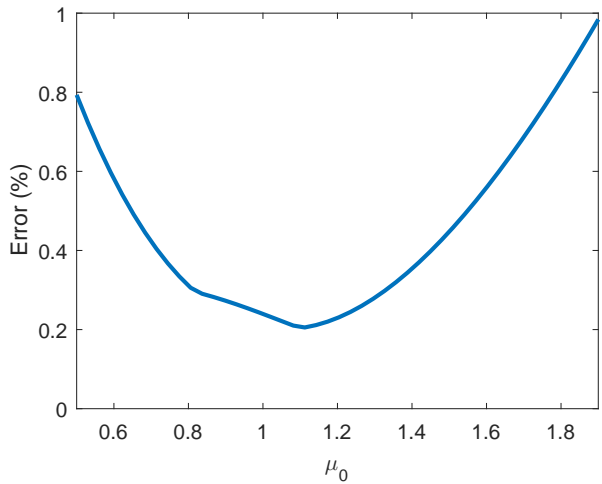}
		\caption{\footnotesize{Errors in $f(z)$ for the phenomenological parametrization (\ref{33d}) for the $\Lambda$CDM background.}}
  \label{Figure_6a}
\end{figure}

%%%%%%%%%%%%%%%%%%%%%%%%%%%%%%%%%%%%%%%%%%%%%%%%%%%%%%%%%%%%%%%%%%%%%%%%%%%%%%%%%%%%
Let us consider the limit $\vert\mu_0-1\vert \ll 1$ in the parametrization given in (\ref{32}).  Using equation (\ref{24b}) we find that,
\begin{eqnarray}
\epsilon(a)&=&\frac{3}{2} \, (\mu_{0}-1) \,\, e^{-g(a)} \, \nonumber \\
&\times&\int_{a_{i}}^{a} \, \Omega_{m}^{1-\gamma^*}(a') \, \frac{1-\Omega_{m}(a')}{1-\Omega_{m}} \,\, e^{\,g(a')} \, \frac{da'}{a'}, 
\end{eqnarray}
and we can fit this expression as follows,
\begin{eqnarray}
\epsilon(a)&=&(\mu_{0}-1) \, (0.505-0.646\,\Omega_m (a)+0.141 \, \Omega_m (a)^{2}). \nonumber\\
\end{eqnarray}
In this case, as we can see in Fig. \ref{Figure_6a}, the error in the growth function is below $1 \%$ for  $\vert\mu_0-1\vert < 1$. Since, as in the $f(R)$ case,  these fits have been obtained 
for a $\Lambda$CDM  background, using the above expressions with different backgrounds would increase the errors up to $2\%$.

%
%%%%%%%%%%%%%%%%%%%%%%%%%%%%%%%%%%%%%%%%%%%%%%%%%%%%%%%%%%%%%%%%%%%%%%%%%%%%%%%%%%%%
\section{Fisher matrix and forecasts}\label{sec7}
%%%%%%%%%%%%%%%%%%%%%%%%%%%%%%%%%%%%%%%%%%%%%%%%%%%%%%%%%%%%%%%%%%%%%%%%%%%%%%%%%%%%
The fitting functions for $f$ in terms of the $\mu$ parameter that 
have been obtained in previous sections can be helpful to 
asses the capability of future galaxy surveys to constrain modified 
gravity theories in a model independent way. Let us consider
 the simplest possibility in which we assume 
that for the range of redshifts and scales  covered by the survey, the
$\mu$ factor can be considered as constant.  In such a case  (\ref{15a})
provides an accurate approximation for the growth function.

The galaxy power spectrum in redshift space, $P(k_{r},\hat{\mu}_{r},z)$ 
is given by \cite{Seo:2003pu}
\begin{equation}\label{6f}
P(k_{r},\hat{\mu}_{r},z)=\frac{D_{A \,r}^{2} \, E(z)}{D_A^{2} \, E_{r}(z)} \, (b+ f  \, \hat{\mu}^{2})^{2} \, D^2{P}(k) \, e^{-k_{r}^{2} \, \hat{\mu}_{r}^{2} \, \sigma_{r}^{2}},
\end{equation}
where, $\hat \mu$ is the angle between $\vec k$ and the line of sight,   $H(z)=H_0E(z)$, $D(z)=\delta_m(z)/\delta_m(0)$ is the growth factor, $b$ is the bias and $f$ is the growth function given by (\ref{15a}) in terms of $\mu$. The sub-index $r$ denotes that the quantity is evaluated 
for the fiducial model.  $\sigma_{r}=\delta z \, (1+z)/H(z)$ with $\delta z(1+z)$ the photometric redshift error and $D_A$ is the angular distance, which, in a flat Universe reads $D_A(z)=(1+z)^{-1} \, \chi (z)$, where $\chi(z)$ is the comoving radial distance,
\begin{equation}\label{3f}
\chi(z)=H_{0}^{-1} \, \int_{0}^{z} \frac{dz'}{E(z')},
\end{equation}
$P(k)$ is the matter power spectrum which is the output of CAMB \cite{Lewis:1999bs} using $\textrm{h/Mpc}$ units. Finally, the dependence $k=k(k_{r})$, $\hat{\mu}=\hat{\mu}(\hat{\mu}_{r})$ and the factor 
$\frac{D^2_{Ar}E}{D^2_{A} \, E_{r}}$ are due to the Alcock-Paczynski effect \cite{Alcock:1979mp},
\begin{equation}\label{7f}
k=Q \, k_{r},
\end{equation}
\begin{equation}\label{8f}
\hat{\mu}=\frac{E \, \hat{\mu}_{r}}{E_{r} \, Q},
\end{equation}
\begin{equation}\label{9f}
Q=\frac{\sqrt{E^{2} \, \chi^{2} \, \hat{\mu}^{2}_{r}-E_{r}^{2} \, \chi_{r}^{2} \, (\hat{\mu}^{2}_{r}-1)}}{E_{r} \, \chi}.
\end{equation}
We consider a simple model in which we have only two free parameters $\mu$ and $\Omega_m$ with a fiducial $\Lambda$CDM cosmology with   $\Omega_{c} \, h^{2}=0.121$, $\Omega_{b} \, h^{2}=0.0226$, $\Omega_{\nu} \, h^{2}=0.00064$, $n_{s}=0.96$, $h=0.68$, $H^{-1}_{0}=2997.9 \, \textrm{Mpc/h}$, $\Omega_{k}=0$ and $\sigma_{8}=0.82$. In this case
\begin{equation}\label{10f}
E(z)=\sqrt{\Omega_{m} \, (1+z)^{3}+(1-\Omega_{m})}.
\end{equation}
 For the bias  we write $b(z)=\sqrt{1+z}$. 
 
Denoting the free model parameters as $p_\alpha=(\Omega_m,\mu)$
 the corresponding Fisher matrix for clustering at a given redshift bin centered at $z$ is \cite{Seo:2003pu},
\begin{widetext}
\begin{equation}\label{1f}
F_{\alpha \beta}=\frac{1}{8 \pi^{2}} \int_{-1}^{1} d\hat{\mu}_r \int_{k_{\text{min}}}^{k_{max}} k_r^{2} \, V_{eff} \, \left.\frac{\partial \ln(P(k_{r},\hat{\mu}_{r},z))}{\partial p_\alpha}\right|_{r} \, \left.\frac{\partial \ln(P(k_{r},\hat{\mu}_{r},z))}{\partial p_\beta}\right|_{r} \, \, dk_r,
\end{equation}
\end{widetext}
where  $V_{eff}=V_{eff}(k,\hat{\mu},z)|_{r}$ is the effective volume, 
\begin{equation}\label{2f}
V_{eff}=\left(\frac{n(z) \, P(k,\hat{\mu},z)}{1+n(z) \, P(k,\hat{\mu},z)}\right)^{2} \, V_{s},
\end{equation}
$n(z)$ is the mean galaxy density at redshift $z$ and $V_{s}=\frac{4 \pi \, f_{sky}}{3} \, \left(\chi(z_{max})^{3}-\chi(z_{min})^{3}\right)$  is the total volume of the survey where $f_{sky}$ is the sky fraction, $z_{max}$ and $z_{min}$ the maximum and minimum redshifts  respectively.  $k_{\text{min}}$ is fixed to 0.007 $h$/Mpc and $k_{max}(z=1.3)=0.218 \,h/$Mpc \cite{Amendola:2013qna}. This value of $k_{max}$ corresponds to $\sigma(z,k_{max})=0.35$ so that only linear
scales are considered in the calculation.

Notice that thanks to the explicit expression in (\ref{15a}), now it is 
possible to compute the derivatives with respect to the 
$\mu$ parameter appearing in the Fisher matrix in a straightforward way. 

As an example, we consider and Euclid-like survey \cite{Euclid} with a unique bin centered at $z = 1.3$ with $z_{min} = 0.5$, $z_{max} = 2.1$, $f_{sky} = 0.375$ which correspond to 15500 $\mathrm{deg}^2$, $\delta z = 0.001$ and $n(z) = 1.12 \times10^{-3}$ $(\textrm{h/Mpc})^{3}$. 

\begin{figure} %[h!]
  	\includegraphics[width=0.4975\textwidth]{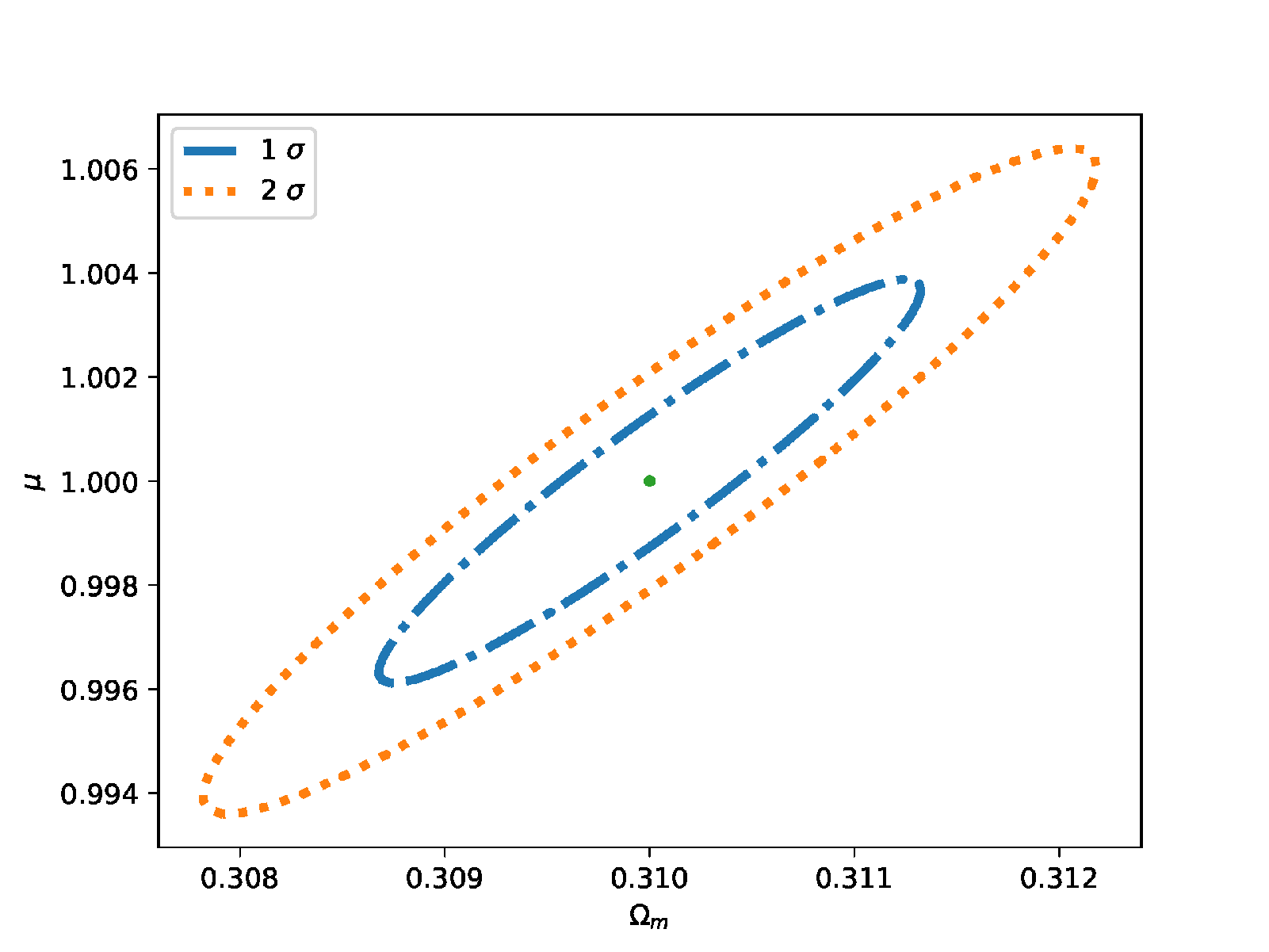}
		\caption{\footnotesize{1-$\sigma$ and 2-$\sigma$ contours for $(\Omega_m, \mu)$ using an example of future galaxy redshift survey with $z$ in $[0.5, 2.1]$, 15500 $\mathrm{deg}^2$, $\delta z = 0.001$ and a mean galaxy density of $n(z) = 1.12 \cdot 10^{-3}$ $(\textrm{h/Mpc})^{3}$}.}
  \label{Figure_7a}
\end{figure}

 Inverting the Fisher matrix, the marginalized error for the $p_\alpha$ parameter is $\sqrt{F^{-1}_{\alpha\alpha}}$. Thus for $\mu$ and $\Omega_m$  we find: $\Delta \Omega_m = 8.76\times 10^{-4}$, $\Delta \mu = 0.0026$, which corresponds to  $\Delta \Omega_m / \Omega_m (\%) = 0.28 \%$, $\Delta \mu / \mu (\%) = 0.26 \%$. In Fig. \ref{Figure_7a}  we plot the 1-$\sigma$ and 2-$\sigma$ contours for $(\Omega_m, \mu)$ 
 assuming that the probability distribution function is Gaussian.

%

%%%%%%%%%%%%%%%%%%%%%%%%%%%%%%%%%%%%%%%%%%%%%%%%%%%%%%%%%%%%%%%%%%%%%%%%%%%%%%%%%%%%
\section{Discussion of results and conclusions}\label{sec6}
%%%%%%%%%%%%%%%%%%%%%%%%%%%%%%%%%%%%%%%%%%%%%%%%%%%%%%%%%%%%%%%%%%%%%%%%%%%%%%%%%%%% 

In this work we aimed at obtaining  fitting formulae for the growth function of 
matter density perturbations in modified gravity theories in a model-independent way.
With that purpose, we have considered the $(\mu, \eta)$ parametrization of modified gravities and shown that a generic expression like $f(a)=\beta(a)\Omega_m^\gamma(a)$
provides good fits to the numerical solutions. 

In the time-independent $\mu(k)$ case,  explicit expressions for $\beta(\mu)$
 and $\gamma(\mu)$ have been obtained for  $\Lambda$CDM and modified backgrounds, yielding accuracies better than $0.5\%$. 

In the time-dependent but scale-independent $\mu(a)$ case, it is not possible to obtain explicit formulae for $\beta$ and $\gamma$, however, it is shown that
a constant $\gamma$ and a $\beta(a)$ function with the same scale factor dependence as
$\mu(a)$ provides errors which again can be in the range of $0.5 \%$.

Finally, in the general case $\mu(a,k)$, it is possible to obtain explicit generic expressions in the perturbative regime $\vert 1-\mu\vert \ll 1$. As an example, for the Hu-Sawicki $f(R)$ model, accuracies below $1\%$ are obtained.

The general perturbative expressions derived in this work exhibit explicit dependence both on the wavenumber $k$ and on the model parameters which could be useful in the 
forecast analysis of modified gravity parameters or in the analysis of growth data of present and future galaxy surveys. A simple Fisher analysis 
for a future Euclid-like survey has been presented as example. Work is in progress in this direction.

\vspace{0.2cm}
{\bf Acknowledgements:} We would like to thank M. Quartin for useful comments that  motivated this work and Eric V. Linder for helpful comments. M.A.R acknowledges support from UCM predoctoral grant.  This work has been supported by the MINECO (Spain) projects FIS2014-52837-P, FIS2016-78859-P(AEI/FEDER, UE),
 and Consolider-Ingenio MULTIDARK CSD2009-00064.

%%%%%%%%%%%%%%%%%%%%%%%%%%%%%%%%%%%%%%%%%%%%%%

\end{document}